

\input harvmac

\Title{BROWN-HET-887}{Some Remarks on Tachyon Action in 2d String Theory}
\centerline{Miao Li\foot{E-mail: li@het.brown.edu}}
\bigskip
\centerline{Department of Physics}
\centerline{Brown University}
\centerline{Providence, RI 02912}
\bigskip
We discuss the effective action of tachyon in the two dimensional string
theory at tree level. We show that already starting from the cubic terms
the action is nonlocal and the usually assumed simplest cubic term does not
give the correct amplitude. Four point 1PI terms are also discussed.

\Date{12/92}

Two dimensional string theory is a very simple string model, compared to
the critical bosonic string and any critical superstring theory. The only
dynamic mode is the tachyon field, besides some topological modes \foot{
It should be noted that the discrete states in the 2d string occur
at imaginary energies, therefore are nonpropagating. Their major role
may be to provide time-dependent backgrounds.}. Yet it
is a nontrivial toy model, for its structure is rich and much of it
is still to be unravelled. In a flat background with a linear dilaton,
perturbation theory is ill-defined without a tachyon background, namely
the cosmological term, since there is a large strong coupling region. Some
even speculates that a tachyon background could be dynamically switched on,
albeit via some nonperturbative effects, in the end of a process such as
Hawking radiation of a blackhole \ref\witten{E. Witten, Phys. Rev. D44
(1991) 314.}. Once a tachyon background is present,
any incident tachyon mode will be prevented from penetrating into the
strong coupling region.

One of remarkable features of the 2d string theory is the absence
of bulk tachyon amplitudes \ref\kel{I. Klebanov, ``String theory
in two dimensions'', Lectures at ICTP Spring School on String Theory and
Quantum Gravity, Trieste, April 1991.}. There is only one delta function
in the scattering amplitude due to the conservation of energy. This is
not surprising, since there is no conservation of momentum because of
the existence of a linear dilaton. The scattering of tachyon is solely
due to the wall effects.
There is only a finite interaction region for tachyon modes with a given
energy, where the coupling constant is strong enough yet not too strong
so that the tachyon modes still can penetrate in the tachyon barrier.

It is conventional to write the following action without a tachyon background
\eqn\act{S=\int\sqrt{-g}e^{-2\Phi}(R+4(\nabla\Phi)^2+(\nabla T)^2 +2T^2
+{a\over 6}T^3+\dots -8),}
where $a$ is a constant and dots denote higher order terms of the tachyon
potential. In the 2d string theory, it is questionable to use such a seemingly
background independent action. There are several points to argue against use
of the above action. First, perturbation theory does not exist for small
fluctuations of $T$. This can be easily seen by rescaling $T$ to absorb the
dilaton factor. There will be extra factor $\hbox{exp}(\Phi)$ in the cubic
term. This is precisely the string coupling constant and there is no bound
of it with a linear dilaton field. Second, equation of motion for the tachyon
derived from \act\ does not alow an exact solution $\hbox{exp}(2\phi)$
whenever $a\ne 0$, where
$\phi$ is the Liouville dimension.
We are convinced by some calculations of tachyon maplitudes on the world sheet
\ref\gl{M. Goulian and M. Li, Phys. Rev. Lett. 66 (1991) 2051; P. Di Francesco
and D. Kutasov, Phys. Lett. B261 (1991) 385.} that such a tachyon background
is exactly marginal, at least in that framework of regularization. Another
perhaps unpleasant fact is that no matter what the exact tachyon background
is, when it is plugged into the equation of motion for the metric derived
from \act, there is backreaction to the metric. The metric is no longer flat
\ref\de{S.P. de Alwis and J. Lykken, Phys. Lett. B269 (1991) 264.}.

So we are forced to introduce a tachyon background into the action.
As will be seen, to reproduce even three point amplitude calculated by
different methods \ref\amp{K. Demeterfi, A. Jevicki and J.P. Rodrigues,
Nucl. Phys. B362, (1991) 173; J. Polchinski, Nucl. Phys. B362 (1991) 125;
G. Moore, Nucl. Phys. B368 (1992) 557; G. Mandal, A.M. Sengupta and
S.R. Wadia, Mod. Phys. Lett. A6 (1991) 1465.}, we have to introduce perhaps
infinitely
many cubic terms. This is a nonlocal action, similar to the nonlocal
quadratic action introduced in \ref\ms{G. Moore and N. Seiberg, Int. Journ.
Mod. Phys. A7 (1992) 2601.}. It is appropriate to point out that we are
considering
the effective action for the tachyon. Although other modes such as the
graviton do not propagate, integrating out these modes in principle
generates nonlocal terms in the tachyon effective action. A similar
situation occurs in the 2d Schwinger model, where integrating out the
gauge field induces nonlocal terms. It is not clear whether the nonlocal
terms we need to introduce can be generated in this way.

Let the dilaton field be $\Phi=2\phi$. The simple generalization of the
kinetic action is
\eqn\free{S_0=\int d\phi dt\left({1\over 2}(\partial_t T)^2-{1\over 2}
(\partial_\phi T)^2- 2\mu
e^{2\phi} T^2\right),}
where $T$ is the re-scaled tachyon field. The effect of the tachyon background
is encoded in the last term. $T=0$ is a solution to the equation
of motion, this means that the tachyon background should be an exact solution
to some unknown background independent equation of motion. We should emphasize
at this point that this background independent action is highly nontrivial
in view of the above effective action with the tachyon background.
The Das-Jevicki collective field theory \ref\antal{S.R. Das and A. Jevicki,
Mod. Phys. Lett. A5 (1990) 1639.} is not such a background independent theory
yet, as the parameter $\mu$ enters explicitly as the lagrangian
multiplier. With \free\ as our quadratic action, we will see that
perturbation is now well-defined. Seiberg and Shenker argue that a tachyon
background represents a superselection sector, so should not be minimized over
\ref\ss{N. Seiberg and S. Shenker, Phys. Rev. D45 (1992) 4581.}. This may
be tied up with the difficulty of writing down a background independent
action.

The equation of motion for $T$ is simply
$$(-\partial_t^2+\partial_\phi^2-4\mu e^{2\phi})T=0.$$
This is just the Wheeler-de Wit equation which is satisfied exactly by
the macroscopic loop \ref\mss{G. Moore, N. Seiberg and M. Staudacher, Nucl.
Phys. B362 (1991) 665.}. The solution which drops to zero when
$\phi\rightarrow \infty$ and satisfies the
asymptotic condition
$$T(\phi, t)\rightarrow e^{i\omega(\phi -t)}+S(\omega)e^{-i\omega(\phi+t)}$$
when $\phi\rightarrow -\infty$ is
\eqn\mod{T_\omega (\phi, t)={1\over\pi}\sqrt{\hbox{sinh}\pi\omega}K_{i\omega}
(2\sqrt{\mu}e^{\phi})e^{-i\omega t}.}
The first term in the asymptotic form of $T$ is the incoming plane wave,
the second is the reflected outgoing plane wave. $S(\omega)$ is just the
reflection coefficient. It is easily found from the definition of function
$K$:
\eqn\refl{S(\omega)=-\mu^{-i\omega}{\Gamma(1+i\omega)\over
\Gamma(1-i\omega)},}
this is the exact tree level two point amplitude, it is a pure phase
and can be absorbed by a redefinition. It is straightforward to
establish the following orthogonal relation
$$\int^\infty_{-\infty}d\phi T_{\omega'}(\phi)T_\omega(\phi)=
{1\over 2 \omega}\delta (\omega - \omega'),$$
where $T_\omega(\phi)$ is the function in \mod\ without the time factor.
It is a real function of $\phi$.

Before setting off to discuss scattering amplitudes, we perform the
standard second quantization. First we should emphasize that for a given
$\omega$, there is only one single particle state represented by \mod\
which contains the incident right moving mode and the reflected left
moving mode. In the usual quantum field theory with one spatial dimension,
there are two modes associated with a given energy $\omega$, namely
the right moving and left moving modes. In our case, if we like, we can view
the incident right moving mode as an in-state, and the reflected mode
as an out-state, then the reflection coefficient \refl\ represents the
time delay. Now we expand the quantum field $T$ in terms of $T_\omega$:
\eqn\exp{T(\phi, t)=\int_0^\infty d\omega (a_\omega T_\omega(\phi, t)
+a_\omega^+ T_\omega^*(\phi, t)),}
where $a_\omega$ and $a_\omega^+$ are destruction and creation operators
respectively. With the help of the orthogonal relation, we obtain the
commutation relation
$$[a_{\omega'}, a_\omega^+]=\delta (\omega'-\omega).$$
The hamiltonian corresponding to the free action \free\ is then diagonalized:
$$H=\int_0^\infty d\omega \omega a^+_\omega a_\omega .$$
There are only half modes as can be seen obviously from the above equation.
We comment at this point that the authors of \ms\ introduced a nonlocal
quadratic action which is significantly different from \free\ at high
energies, as they believe this is the signal of soft high energy behavior
of a string model. This nonlocality is certainly the most common feature
in the model under discussion, as it arises also in the cubic interaction
as will be seen soon. If we use the nonlocal quadratic action in \ms,
the three point amplitude calculated from the simplest cubic action is even
more different from the correct answer. We will not use that quadratic
action here.

The most natural cubic term in the lagrangian is proportional to
$\hbox{exp}(2\phi)T^3$, here the exponential factor represents the
string coupling constant. It is easy to discuss three point amplitudes
in the Hamiltonian framework. We then write the contribution of this
term to the interaction part of hamiltonian as
\eqn\inte{H_I^{(3)}= -{a\over 3!} \int d\phi e^{2\phi}:T^3:,}
where $:\quad :$ denotes the normal ordering, $a$ is a constant to be
determined.

Three point amplitude $T(\omega_1\rightarrow \omega_1'+\omega_2')$ will be
calculated only, since the result for another amplitude is the same. The
correct answer gievn in \amp\ is
\eqn\corr{ T(\omega_1\rightarrow \omega_1'+\omega_2')= {2\pi i\over 8\pi \mu}
\delta(\omega_1-\omega_1'-\omega_2')\sqrt{\omega_1\omega_1'\omega_2'},}
a fairly simple result.

Now expanding
$H_I^{(3)}$ in terms of modes, the relevant part for this amplitude is
$$-{a\over 2}\int d\phi e^{2\phi}(\int d\omega_1d\omega_2d\omega_3
T^*_{\omega_1}T^*_{\omega_2}T_{\omega_3}a^+_{\omega_1}a^+_{\omega_2}
a_{\omega_3} ).$$
The three point amplitude is read off straightforwardly from the above equation
\eqn\ampl{\eqalign{T_0(\omega_1\rightarrow \omega_1'+\omega_2')=2\pi ai\delta
(\omega_1-\omega'_1-\omega'_2)f_{\omega_1'}f_{\omega_2'}f_{\omega_1}&\cr
\int_{-\infty}^\infty d\phi e^{2\phi}K_{i\omega_1}(2\sqrt{\mu}e^\phi)K_{
i\omega_1'}(2\sqrt{\mu}e^\phi)K_{i\omega_2'}(2\sqrt{\mu}e^\phi)&,}}
where
$$f_\omega={1\over\pi}\sqrt{\hbox{sinh}\pi\omega}.$$
The reason for us to denote this amplitude by $T_0$ is that this does not
give us the correct one and we reserve the name $T$ for the correct one.
The low energy behavior of the integral involving  Bessel functions in
\ampl\ is discussed in \ss\ and agrees with calculations done in the
matrix model approach. We shall perform the integral exactly and show that
the amplitude differs significantly from the standard result \corr\ at high
energies. This will force us to introduce more cubic terms in
order to reproduce the correct amplitude. Using new variable $x=\hbox{exp}
(\phi)$, the integral reads
$${1\over (2\sqrt{\mu})^2}\int_0^\infty dxxK_{i\omega_1}(x)K_{i\omega_1'}(x)
K_{i\omega_2'}(x).$$
This integral is well defined, as the strong coupling region is suppressed
by the wave functions. We also see explicitly how the effective coupling
constant $1/\mu$ emerges. We are interested in the case when $\omega_1
=\omega_1'+\omega_2'$, the integral is performed in the appendix in this case.
The result is
$${\pi\over 8\mu\hbox{sinh}(\pi\omega_1)}\hbox{Im}\left(\sum_{n=0}^\infty
B(1-i\omega_1'+n, 1-i\omega_2'+n)\right).$$
The sum of beta functions can be expressed in terms of a generalized
hypergeometric function. It seems unlikely that this can be reduced
to a combination of elementary functions. To estimate the high energy behavior,
we write the sum of beta functions as an intgeral
$$\int_0^1 dtt^{-i\omega_1'}(1-t)^{-i\omega_2'}{1\over 1-t+t^2}.$$
When both $\omega_1'$ and $\omega_2'$ are small, the imaginary part of this
integral is about
$$-\omega_1\int_0^1 dt {\hbox{ln}t\over
1-t+t^2}=C\omega_1,$$
where the constant $C=1/3(\psi'(1/3)-2\pi^2/3)=0.781302\dots$. Subsituting
this back to \ampl\ we find that the low
energy behavior of the scattering amplitude agrees with the correct one,
provided the constant in the cubic action is
$$a={\pi^{3/2}\over C}$$
When $\omega_1'$ and $\omega_2'$ are both large, the integral can be estimated
by the stationary phase approximation. Although the amplitude of this integral
is about the same as one should expect from the correct scattering amplitude,
it oscillates fast. The high energy behavior of the scattering amplitude
calculated from the cubic action is then
\eqn\osc{\eqalign{T_0(\omega_1\rightarrow \omega_1'+\omega_2')&=2\pi a i\delta
(\omega_1-\omega_1'-\omega_2')f_{\omega_1}f_{\omega_1'}f_{\omega_2'}\cr
& {\pi\over 8\mu\hbox{sinh}\pi\omega_1}{\sqrt{2\pi\omega_1\omega_1'\omega_2'}
\over\omega_1^2-\omega_1'\omega_2'}\hbox{sin}\left(\hbox{ln}
{\omega_1^{\omega_1}\over(\omega_1')^{\omega_1'}(\omega_2')^{\omega_2'}}
+{\pi\over 4}\right).}}

We have seen that the simplest cubic term does not generate the correct three
point amplitude, although it does at low energy. Apparently there are many
ways to introduce additional cubic terms in order to generate the three point
amplitude. We give one simple choice, maybe not the minimal one, in the
following. Let $f(\omega_1',\omega_2')=T/T_0$, the ratio of the correct
amplitude and the amplitude generated by the simple cubic term. This function
has an expansion
$$f(\omega_1',\omega_2')=\sum_{m, n=0}f_{mn}(\omega_1')^m(\omega_2')^n.$$
The leading term is $1$. Consider a function $g(\omega_1',\omega_2')$ satisfy
$${1\over 3}(g(\omega_1',\omega_2')+g(\omega_1'+\omega_2',\omega_1')
+g(\omega_1'+\omega_2',\omega_2'))=f(\omega_1',\omega_2').$$
Given a solution, expanding it into power series
$$g(\omega_1',\omega_2')=\sum_{m,n=0}g_{mn}(\omega_1')^m(\omega_2')^n,$$
we introduce in the action the cubic terms
\eqn\nonl{{a\over 3!}\sum_{m,n=0}g_{mn}e^{2\phi}T(H^{m/2}T)(H^{n/2}T),}
where operator $H$ is
$$H=-{\partial^2\over \partial\phi^2}+4\mu e^{2\phi}.$$
\nonl\ will reproduce the correct three point amplitude, obviously it is
a nonlocal action.

Four point amplitudes are more difficult to calculate. Consider for example
amplitude $T(\omega_1+\omega_2\rightarrow \omega_1'+\omega_2')$. There are two
difficulties. First, to calculate the one point reducible part, we have
to know the off-shell three point amplitude, namely when $\omega_1\ne
\omega_1'+\omega_2'$ in \ampl. Second, to calculate the 1PI part, we need
to calculate an integral involving the product of four Bessel functions.
Let us first discuss the one point reducible part. According to the standard
formula, this amplitude is given by
\eqn\red{2\pi i\delta(\omega_1+\omega_2-\omega_1'-\omega_2')\int_0^\infty
d\omega_3{T(\omega_3\rightarrow \omega_1+\omega_2)T(\omega_3\rightarrow
\omega_1'+\omega_2')\over \omega_1+\omega_2-\omega_3-i\epsilon} ,}
where $T(\omega_3\rightarrow \omega_1+\omega_2)$ is the three point amplitude
containing no delta function corresponding to energy conservation, $\epsilon$
is a small positive constant. The off-shell three point amplitude depends
crucially on what cubic action we choose, as we know there is no unique way
to choose one to reproduce the correct on-shell amplitude. According to our
choice made before, the off-shell amplitude is
$$T(\omega_3\rightarrow\omega_1+\omega_2)=T_0(\omega_3\rightarrow \omega_1+
\omega_2)f(\omega_1,\omega_2),$$
where $T_0$ (without the delta function factor) is calculated by the
simplest cubic action $\hbox{exp}(2\phi) T^3$ with arbitrary $\omega_i$,
and $f(\omega_1,\omega_2)$ is the function defined before. A formula for $T_0$
is given in the appendix.

Now we turn to calculating the 1PI amplitude, starting from
$$H_I^{(4)}=-{b\over 4!}\int d\phi e^{4\phi}:T^4:,$$
where $b$ is another constant. Denote the amplitude calculated from this
hamiltoian by $T_0$ again,
\eqn\four{\eqalign{&T_0(\omega_1+\omega_2\rightarrow\omega_1'+\omega_2')
=2\pi bi\delta(
\omega_1+\omega_2-\omega_1'-\omega_2')f_{\omega_1}f_{\omega_2}f_{\omega_1'}
f_{\omega_2'}\cr
&\int_{-\infty}^\infty d\phi e^{4\phi}K_{i\omega_1}(2\sqrt{\mu}e^\phi)
K_{i\omega_2}(2\sqrt{\mu}e^\phi)K_{i\omega_1'}(2\sqrt{\mu}e^\phi)K_{i\omega_2'}
(2\sqrt{\mu}e^\phi).}}
Again an integral with Bessel functions need be calculated. Using $x=e^\phi$,
the integral in \four\ is just
\eqn\for{{1\over (2\sqrt{\mu})^4}\int dx x^3K_{i\omega_1}(x)K_{i\omega_2}
(x)K_{i\omega_1'}(x)K_{i\omega_2'}(x),}
where we see how the coupling constant $1/\mu$ appears naturally. We have
tried to reduce this integral to a simpler one. It turns out that this is
possible, but the simpler integral can not be carried out. To estimate the
low energy behavior of \for, use the following transform
$$K_{i\omega}(x)=\int dt e^{-x\hbox{cosh}t}\hbox{cos}(\omega t).$$
{}From this formula we immediately see that when all $\omega$'s are small, the
leading term in \for\ is a constant. So \four\ does give the right low energy
behavior. The next order of the integral \for\ is quadratic in $\omega$'s,
therefore \four\ alone does not give the correct next order term \amp.

It is not particularly illuminating to exactly calculate the one point
reducible part in \red\ and the irreducible part in \four, and to combine
them to get a closed formula. Suffices it to say that the simplest quartic
term together with the cubic terms we have introduced does not give the
correct four point amplitude, as given in \amp. Once again infinitely many
quartic terms are to be introduced.

In conclusion we have seen that to reproduce scattering amplitudes calculated
by other means, infinitely many terms have to be introduced in the effective
action of tachyon. This effective action is nonlocal, and we suspect that
nonlocality persists to all orders. We have also noted that the effective
action can not be fixed uniquely, upon comparing scattering amplitudes. It may
be helpful to employ a spacetime realization of those discrete symmetries
\ref\sym{J. Avan and A. Jevicki, Phys. Lett. B266 (1991) 35; B272 (1991) 17;
D. Minic, J. Polchinski and Z. Yang, Nucl. Phys. B369 (1992) 324; S. Das,
A. Dhar, G. Mandal and S. Wadia, Mod. Phys. Lett. A7 (1992) 71.} \ref\ward{
I.R. Klebanov, Mod. Phys. Lett. A7 (1992) 723; I.R. Klebanov and A.
Pasquinucci, preprint PUPT-1313; A. Jevicki, J.P. Rodrigues and A.J. van
Tonder, preprint BROWN-HET-874.} to fix the effective action.
In view of the above discussion, it is remarkable
that the Das-Jevicki collective field theory summarizes the physical content
neatly in a cubic action, with derivative couplings. Discrete symmtries are
also readily realized in that theory. There must be an interesting
transformation between the usual effective action and the collective field
theory. This transformation is most desirable to know in order to better
understand spacetime physics of the 2d string, and eventually to uncover
a background independent formulation. Finally, we comment that it is not
clear whether the collective field theory is just the effective theory
for tachyon in the flat background, or it indeed encodes more information
about the 2d string. To go beyond the flat background such as to discuss
blackhole physics in this framework therefore requires more justification.

\vskip1cm
\noindent {\bf Acknowledgements}

I am grateful to S. Giddings for discussions concerning tachyon potential
occurred between us about two years ago. I wish to thank A. Jevicki for
many helpful discussions. This work was supported by DOE contract
DE-FG02-91ER40688-Task A.

\appendix{A}{}

In this short appendix we calculate integral
\eqn\bess{\int dxxK_{i\omega_1}(x)K_{i\omega_1'}(x)K_{i\omega_2'}(x).}
Use the integral representation of the product of two Bessel functions
\ref\tabl{I.S. Gradshteyn and I.M. Ryzhik, Table of Integrals, Series,
and Products, Academic Press Inc. (1980).}
$$K_{i\omega_1'}(x)K_{i\omega_2'}(x)=2\int_0^\infty dt\hbox{cos}(\omega_1'
-\omega_2')tK_{i(\omega_1+\omega_2)}(2x\hbox{cosh}t)$$
the integral \bess\ is reduced to
$$2\int_0^\infty dt\hbox{cos}(\omega_1'-\omega_2')t\int_0^\infty dxx
K_{i\omega_1}(x)K_{i(\omega_1'+\omega_2')}(2x\hbox{cosh}t)$$
The integral involving the product of two Bessel functions can be expressed
in terms of the hypergeomtric function \tabl. When $\omega_1=\omega_1'
+\omega_2'$, this hypergeometric function reduces to an elementary function,
so \bess\ reduces to
\eqn\res{\eqalign{&{\pi\over i\hbox{sinh}\pi\omega_1}\int_0^\infty dt
{\hbox{cos}(\omega_1'-\omega_2')t\over 4\hbox{cosh}^2t-1}\left((2\hbox{cosh}
t)^{i\omega_1}-(2\hbox{cosh}t)^{-i\omega_1}\right)\cr
&={\pi\over 2\hbox{sinh}\pi\omega_1}\hbox{Im}\left(\sum_{n=0}^\infty
B(1-i\omega_1'+n,1-i\omega_2'+n)\right).}}

For $\omega_1\ne \omega_1'+\omega_2'$, integral \bess\ can be calculated
similarly, the result is
\eqn\any{\eqalign{&{\pi\over 2\hbox{sinh}\pi\omega_1}\hbox{Im}[\sum_{n=0}
^\infty{\Gamma(1+{i\over 2}
(\omega_1'+\omega_2'-\omega_1)+n)\Gamma(1-{i\over 2}(\omega_1+\omega_1'+
\omega_2')+n)\over n!\Gamma(1-i\omega_1+n)\Gamma(2-i\omega_1+2n)}\cr
&\times\Gamma(1-{i\over 2}(\omega_1+\omega_1'-\omega_2')+n)\Gamma
(1-{i\over 2}(\omega_1+\omega_2'-\omega_1')+n)].}}
Substituting this back into \ampl\ and dropping out the delta function, the
off-shell three point amplitude is obtained.

\listrefs\end